\documentclass[amsmath,amssymb,amsfonts,twocolumn]{revtex4}
\usepackage{graphicx}
\usepackage{bbm}

\def \beq {\begin{equation}}
\def \eeq {\end{equation}}
\def \tr {\rm Tr}

\begin{document}
\title{Retrodictive derivation of the radical-ion-pair master equation and Monte-Carlo simulation with single-molecule quantum trajectories} 
\author{M. Kritsotakis and I. K. Kominis}
\email{ikominis@physics.uoc.gr}

\affiliation{Department of Physics, University of Crete, Heraklion 71103, Greece}

\begin{abstract}
Radical-ion-pair reactions, central in photosynthesis and the avian magnetic compass mechanism, have been recently shown to be a paradigm system for applying quantum information science in a biochemical setting. 
The fundamental quantum master equation describing radical-ion-pair reactions is still under debate. We here use quantum retrodiction to formally refine the theory put forward in Phys. Rev. E {\bf 83}, 056118 (2011). We also provide a rigorous analysis of the measure of singlet-triplet coherence required for deriving the radical-pair master equation. A Monte-Carlo simulation with single-molecule quantum trajectories supports the self-consistency of our approach.
\end{abstract}
\maketitle
\section{Introduction}
Radical-ion pairs and their spin-dependent reactions \cite{steiner,steiner2} have been recently shown \cite{komPRE2009,komPRE2011,komPRE2012,katsop,komCPL,dellis1,dellis2,cidnp,JH,briegel,vedral,plenioPRA,sun,horePRL,plenioPRL,shao,kais} to be a paradigm system for the emerging field of quantum biology \cite{plenio_review}, that is, the study of quantum coherence effects, or in general the study of quantum information science in the context of biological systems. The biological significance of radical-ion-pair (RP) reactions is twofold, (i) they are understood to underlie the avian magnetic compass mechanism  \cite{schulten,ritz,ww,horeNature,rodgers,mouritsen}, and (ii) they participate in the electron-transfer cascade reactions taking place in photosynthetic reactions centers  \cite{boxer,matysik}. In any case, the experimentally founded science of spin-chemistry \cite{woodward} deals with such reactions in a wide range of chemical contexts. Hence the theoretical understanding of RP reactions at the fundamental level is of importance for current experimental work in spin chemistry, for further exploring quantum effects in biological systems as well as for the design of novel, and potentially quantum-limited biomimetic devices and sensors.

Theoretically, the fate of radical-ion-pair reactions and all relevant predictions are fully accounted for by the time evolution of $\rho$, the RP's spin density matrix. The time evolution of $\rho$ was until recently understood to be driven by (i) unitary Hamiltonian evolution due to all magnetic interactions within the RP, and (ii) RP population loss due to spin-dependent charge recombination.  We have recently shown that the spin degrees of freedom of the RP form an open quantum system, i.e. there is a third source of time evolution: (iii) the spin decoherence inherent in the radical-pair mechanism \cite{komPRE2009,komPRE2011}. Moreover, since the RP is in general in a coherent (or partially coherent) superposition of spin states (we refer in particular to singlet-triplet coherence), the description of the RP's reaction kinetics appears not to be as straightforward as originally thought. In \cite{komPRE2011} we demonstrated that singlet-triplet (S-T) coherence of the RP is a central concept in understanding the intimately related effects (i)-(iii) and put forward a master equation satisfied by the density matrix $\rho$. While S-T decoherence was described \cite{komPRE2009} by first-principles perturbation theory (similar to most applications of the theory of Markovian open quantum systems leading to a Lindblad decoherence term), the reaction kinetics had been accounted for in a phenomenological manner open to criticism. Moreover, the introduction \cite{komPRE2011} of the coherence measure $p_{\rm coh}$ quantifying the "strength" of S-T coherence was also done intuitively. 

In this work we formalize our approach along both fronts previously mentioned. In particular, (i) we show that the measure of S-T coherence introduced in \cite{komPRE2011} is not well-defined. We then introduce a new measure of S-T coherence based on recently appeared rigorous considerations by Plenio and co-workers \cite{plenio_coherence}, (ii) we formally derive the reaction terms of the master equation using quantum retrodiction, a concept borrowed from the field of quantum communications, and  (iii) we introduce Monte Carlo (MC) simulation of single-RP quantum trajectories \cite{molmer,wiseman}. The MC simulation contains by design all relevant phenomena at the single-molecule level, and hence forms a unique tool to test the predictions of our master equation. 

We show that the new measure of S-T coherence, properly scaling with the off-diagonal elements of the density matrix, is essential for the decomposition of $\rho$ into a mixture of maximally coherent and maximally incoherent states.
This decomposition underlies the retrodictive derivation of the new reaction terms, which lead to (a) a significantly improved agreement of the new master equation prediction with MC, and 
(b) the derivation of precise and experimentally measurable decay rates for the S-T coherence.

In particular, in Section III we introduce the Monte Carlo simulation of single-RP quantum trajectories including only S-T decoherence and compare it with the master equation for non-recombining RPs where perfect agreement is expected by definition. In Section IV we elaborate on the shortcomings of our previous measure of S-T coherence and then introduce a new measure based on \cite{plenio_coherence}. The decomposition of $\rho$ into a mixture of maximally coherent and maximally incoherent states is presented in Section V. This decomposition is the basis of the rigorous theory of quantum retrodiction used to derive the reaction terms of the master equation, presented in Section VI. In Section VII we perform a Monte Carlo simulation of RP quantum trajectories including recombination, comparing the trajectory-average with the prediction of our new master equation. Finally, in Section VIII we discuss the decay of S-T coherence in a way that could be relevant to experimentally accessible observables and we compare our theory with the predictions of competing theoretical approaches. In the following Section we start with a few definitions and a brief review of previous work in order to make this work as comprehensive as possible for the general reader. 
\section{Definitions and previous work}
The quantum degrees of freedom of RPs are formed by a multi-spin system embedded in a biomolecule. In particular, RPs are biomolecular ions created by a charge transfer from a photo-excited D$^*$A donor-acceptor biomolecular dyad DA, schematically described by the reaction ${\rm DA}\rightarrow {\rm D^{*}A}\rightarrow {\rm D}^{\bullet +}{\rm A}^{\bullet -}$, where the two dots represent the two unpaired electrons of the two radicals. The excited state D$^*$A is usually a spin zero state, hence the initial spin state of the two unpaired electrons is a singlet, denoted by $^{\rm S}{\rm D}^{\bullet +}{\rm A}^{\bullet -}$.

\begin{figure}
\includegraphics[width=5.5 cm]{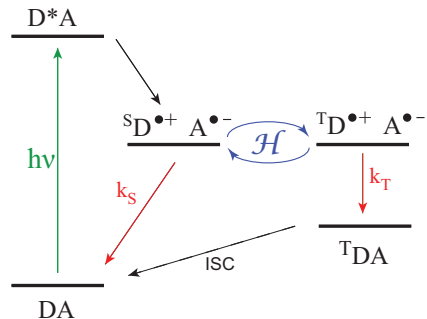}
\caption{(Color online) Simplified energy level diagram depicting radical-ion-pair reaction dynamics. A donor-acceptor dyad is photo-excited and a subsequent charge transfer produces a singlet radical-ion pair. Magnetic interactions within the radical pair induce coherent singlet-triplet mixing, while spin-dependent charge recombination leads to singlet and triplet neutral products at the respective reaction rates $k_{\rm S}$ and $k_{\rm T}$. The reaction can in principle close through intersystem crossing from the triplet to the singlet ground state.}
\label{fig1}
\end{figure}

Now, both D and A contain a number of  magnetic nuclei which hyperfine-couple to the donor's and acceptor's electron, respectively, effectively creating a different magnetic environment for the two unpaired electrons. This leads to S-T mixing, i.e. a coherent oscillation of the spin state of the electrons. Charge recombination terminates the reaction and leads to the formation of the neutral reaction products. Angular momentum conservation at this step empowers the molecule's spin degrees of freedom and their minuscule (relative to thermal) energy to determine the reaction's fate:  singlet state RPs, $^{\rm S}{\rm D}^{\bullet +}{\rm A}^{\bullet -}$, recombine to reform the neutral spin zero DA molecules, whereas triplet RPs, $^{\rm T}{\rm D}^{\bullet +}{\rm A}^{\bullet -}$, recombine to a different (metastable) triplet neutral product $^{\rm T}$DA. For completeness we note that the reaction can, in principle, close through the so-called intersystem crossing $^{\rm T}{\rm DA}\rightarrow {\rm DA}$. The above are schematically shown in Fig. \ref{fig1}. 

The straightforward part of RP dynamics are the unitary dynamics embodied in the magnetic Hamiltonian ${\cal H}$, which mainly contains (i) hyperfine couplings of the donor's (acceptor's) electron with the donor's (acceptor's) nuclear spins, (ii) Zeeman interaction of the donor's and acceptor's electrons with the externally applied magnetic field (nuclear Zeeman interaction is usually neglected), (iii) spin-exchange and dipolar interactions between the donor's and the acceptor's electron \cite{efimova,dellis1}. 

Were this a closed system, its dynamics would be fully described by Liouville's equation $d\rho/dt=-i[{\cal H},\rho]$. However, it is not, hence there are more terms that make up the master equation, and they will be elaborated in the following. These terms involve two central operators, the singlet and triplet projectors ${\rm {\rm Q_S}}$ and ${\rm Q_T}$, respectively. Before defining them, we note that the density matrix $\rho$ describes the spin state of the RP's two electrons and $M$ magnetic nuclei located in D and A. The dimension of $\rho$ is $d=4\Pi_{j=1}^{M}(2I_{j}+1)$, where  $I_j$ is the nuclear spin of the $j$-th nucleus, with $j=1, 2,..., M$. For our numerical work we consider the simplest possible RP, namely an RP containing just one spin-1/2 nuclear spin hyperfine coupled to e.g. the donor's electron. In this case the density matrix has dimension $d=8$. This simple model system exhibits the essential physics without the additional complication of more nuclear spins. We stress that the master equation we derive is general and equally applicable for any number of nuclear spins entering the magnetic Hamiltonian ${\cal H}$ and any sort of interactions included in ${\cal H}$.

Angular momentum conservation at the recombination process splits the RP's Hilbert space into an electron singlet and an electron triplet subspace, defined by the respective projectors ${\rm {\rm Q_S}}$ and ${\rm Q_T}$. These are $d\times d$ matrices given by ${\rm {\rm Q_S}}={1\over 4}\mathbbmtt{1}_{d}-\mathbf{s}_{D}\cdot\mathbf{s}_{A}$ and ${\rm Q_T}={3\over 4}\mathbbmtt{1}_{d}+\mathbf{s}_{D}\cdot\mathbf{s}_{A}$, where $\mathbf{s}_{D}$ and $\mathbf{s}_{A}$ are the spin operators of the donor and acceptor electrons written as $d$-dimensional operators, e.g. the $j$-th component of $\mathbf{s}_{D}$ is written as $s_{jD}=\hat{s}_{j}\otimes\mathbbmtt{1_2}\otimes\mathbbmtt{1}_{2I_1+1}\otimes\mathbbmtt{1}_{2I_2+1}...\otimes\mathbbmtt{1}_{2I_M+1}$, where the first operator in the previous Kronecker product refers to the donor's electron spin, the second to the acceptor's electron spin and the rest to the nuclear spins. By $\hat{\mathbf{s}}$ we have denoted the regular (2 dimensional) spin-1/2 operators and by $\mathbbmtt{1}_{m}$ the $m$-dimensional unit matrix. We note that the RP's singlet subspace has dimension $\Pi_{j=1}^{M}(2I_{j}+1)$ while the triplet subspace has dimension $3\Pi_{j=1}^{M}(2I_{j}+1)$. The electron multiplicity 1 in the former corresponds to the singlet state $|{\rm S}\rangle=(|\uparrow\downarrow\rangle-|\downarrow\uparrow\rangle)/\sqrt{2}$, while the multiplicity of 3 in the latter stems from the three triplet states $|{\rm T}_0\rangle=(|\uparrow\downarrow\rangle+|\downarrow\uparrow\rangle)/\sqrt{2}$, $|{\rm T}_{+}\rangle=|\uparrow\uparrow\rangle$ and $|{\rm T}_{-}\rangle=|\downarrow\downarrow\rangle$.

The projectors ${\rm Q_S}$ and ${\rm Q_T}$ are complete and orthogonal, i.e. ${\rm Q_S+Q_T}=\mathbbmtt{1}_{d}$ and ${\rm Q_S}{\rm Q_T}={\rm Q_T}{\rm Q_S}=0$. There are also two rates to consider, the singlet and triplet recombination rates, $k_{\rm S}$ and $k_{\rm T}$, respectively. These are defined as follows: consider an RP ensemble with no magnetic interactions (${\cal H}=0$) to be in the singlet (triplet) state. Then its population would decay exponentially at the rate $k_{\rm S}$ ($k_{\rm T}$). Finally, in any given time interval $dt$, the measured singlet and triplet neutral products will be $dn_{\rm S}=k_{\rm S}dt\tr\{\rho {\rm {\rm Q_S}}\}$ and $dn_{\rm T}=k_{\rm T}dt\tr\{\rho {\rm Q_T}\}$. These relations are simple to understand, namely in the time interval $dt$ there would be $k_{\rm S}dt$ singlet and $k_{\rm T}dt$ triplet recombinations if all RPs were in the singlet or triplet state, respectively. If they are in the general state described by $\rho$, then $k_{\rm S}dt$ and $k_{\rm T}dt$ have to be multiplied by the respective probabilities to be in the singlet or triplet state. 

The initial state most often considered when doing calculations with the density matrix is the singlet electron-unpolarized nuclear spin state written as $\rho={\rm Q_S}/\tr\{\rm Q_S\}$. 
\subsection{Singlet-Triplet decoherence}
A more detailed look at the energy level structure of Fig. 1 reveals the picture depicted in Fig. 2, where we show the vibrational excited states of the singlet and triplet ground states, which form the singlet and triplet reservoir. Radical-pair recombination proceeds as a {\it real} transition of the RP to one of the quasi-resonant and quasi-continuous reservoir states. As we have demonstrated in \cite{komPRE2012}, there cannot be any coherence between the RP state and the neutral ground states, but only population transfer from the former to the latter, due to which the RP is an open system. What we have shown in \cite{komPRE2009} is that it is "doubly-open", because the same reservoir states lead to S-T decoherence. Using 2$^{\rm nd}$-order perturbation theory we have shown that {\it virtual} transitions to these vibrational reservoir states {\it and back} interrupt the coherent S-T mixing in individual RPs and hence cause the decay of the ensemble S-T coherence. This is described with a Lindblad-type and trace-preserving master equation
\beq
{{d\rho}\over {dt}}\Big|_{\rm decoh}=-i[{\cal H},\rho]-{{k_{\rm S}+k_{\rm T}}\over 2}\big({\rm {\rm Q_S}}\rho+\rho {\rm {\rm Q_S}}-2{\rm {\rm Q_S}}\rho {\rm {\rm Q_S}}\big)\label{MEnr}
\eeq
In other words, this equation describes the null quantum measurement of the RP's neutral reaction products: there is a certain probability that the RP will recombine during a time interval $dt$. If this 
does not happen, i.e. if no reaction product is detected, then there are three different possibilities that could be realized within $dt$, (i) a projection to the singlet state, (ii) a projection to the triplet state and (iii) Hamiltonian evolution. In the following Section we present a Monte Carlo simulation of individual quantum trajectories and elaborate in detail on these issues. 
\begin{figure}
\includegraphics[width=8.5 cm]{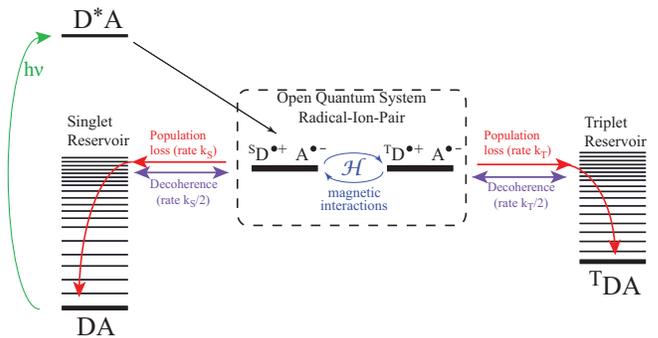}
\caption{(Color online) Detailed energy level structure of radical-ion pairs. The vibrational excitations of the singlet (DA) and the triplet ($^{\rm T}$DA) ground state form a reservoir that probes the electron spin state of the RP, leading to an intramolecule measurement of ${\rm Q_S}$. Virtual transitions (rates $k_{\rm S}/2$ and $k_{\rm T}/2$) to the reservoir levels and back to the RP lead to S-T decoherence, while real transitions (rates $k_{\rm S}$ and $k_{\rm T}$) to the reservoir states followed by their decay to the ground state lead to recombination.}
\label{fig2}
\end{figure}
\begin{figure}
\includegraphics[width=8.5 cm]{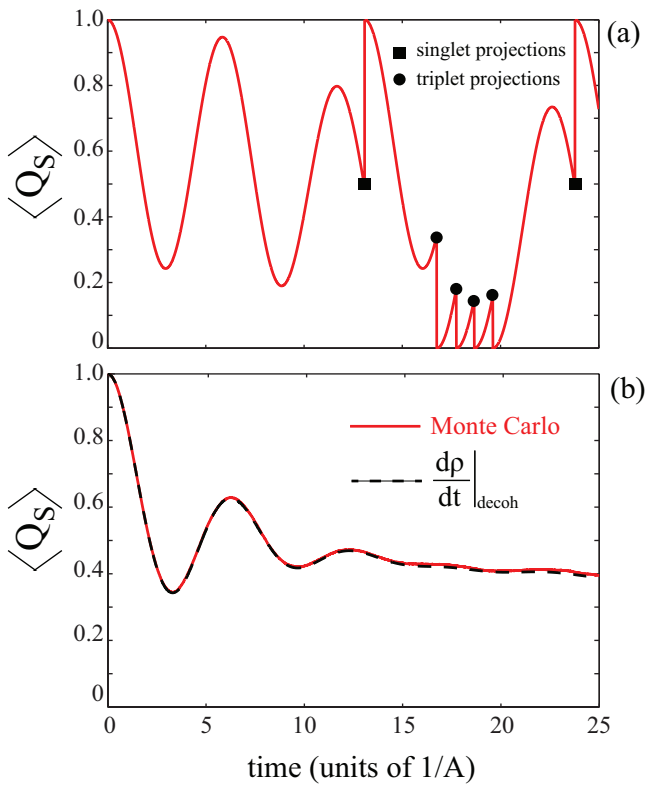}
\caption{(Color online) The time evolution of $\langle{\rm Q_S}\rangle$ for a model RP with one nuclear spin, taking into account only S-T decoherence and S-T mixing driven by the Hamiltonian ${\cal H}=\omega(s_{1z}+s_{2z})+A\mathbf{s}_{1}\cdot\mathbf{I}$, where the Larmor frequency is taken $\omega=A/10$ and the recombination rates are $k_{\rm S}=k_{\rm T}=A/4$. These parameters represent a typical RP at earth's field with a hyperfine coupling on the order of 1 mT and recombination times on the order of 20 ns. (a) single-RP quantum trajectory, depicting singlet and triplet projections at random instants in time. The initial RP state for this trajectory is $|S\rangle\otimes|\uparrow\rangle$. (b) average of 20,000 such trajectories (red solid line), half of which have initial state $|S\rangle\otimes|\uparrow\rangle$ while the other half have initial state $|S\rangle\otimes|\downarrow\rangle$. The time axis was split into 10,000 steps $dt$, in everyone of which one out of the three possibilities outlined in Section III was realized. The prediction of the trace-preserving master equation \eqref{MEnr} is shown by the black dashed line. The initial state for the density matrix was the usually considered singlet state with unpolarized nuclear spin, $\rho={\rm Q_S}/\tr\{\rm Q_S\}$.}
\label{fig3}
\end{figure}
\section{Monte Carlo simulation of S-T decoherence using single-molecule quantum trajectories}
As well known from quantum optics, the absence of a detection event, e.g. a photon detection, in a quantum measurement, called "null" measurement, also has an effect on the system's quantum state. What we have shown in \cite{komPRE2009} is that the quantum state evolution of a non-recombining RP (absence of detection of recombination events) is given by the Lindblad master equation \eqref{MEnr}. This trace-preserving master equation encompasses the following three possibilities a non-recombining RP faces during the time evolution of its quantum state: \newline
{\bf (i)} a quantum jump to the singlet state $\rho_{\rm S}={\rm {\rm Q_S}}\rho {\rm {\rm Q_S}}/\tr\{\rho {\rm {\rm Q_S}}\}$, taking place with probability 
\beq
dp_{\rm S}={{(k_{\rm S}+k_{\rm T})dt}\over 2}\tr\{\rho {\rm {\rm Q_S}}\}
\eeq
{\bf (ii)} a quantum jump to the triplet state $\rho_{\rm T}={\rm Q_T}\rho {\rm Q_T}/\tr\{\rho {\rm Q_T}\}$, taking place with probability 
\beq
dp_{\rm T}={{(k_{\rm S}+k_{\rm T})dt}\over 2}\tr\{\rho {\rm {\rm Q_T}}\}
\eeq
{\bf (iii)} unitary evolution driven by the Hamiltonian ${\cal H}$,  taking place with probability $1-dp_{\rm S}-dp_{\rm T}$.

In an ensemble of RPs, these single-molecule possibilities are unobservable, so we have to average over them.
This averaging {\it exactly} reproduces the master equation \eqref{MEnr}. In other words, writing $\rho_{t+dt}=dp_{\rm S}\rho_{\rm S}+dp_{\rm T}\rho_{\rm T}+(1-dp_{\rm S}-dp_{\rm T})(\rho_t-idt[{\cal H},\rho_t])$ leads to \eqref{MEnr} for $d\rho/dt=(\rho_{t+dt}-\rho_t)/dt$. 

The physical significance of the sum $k_{\rm S}+k_{\rm T}$ appearing in the probabilities $dp_{\rm S}$ and $dp_{\rm T}$ is the fact that both singlet and triplet reservoirs continuously "measure" the same observable, namely ${\rm Q_S}$. The result of this measurement is either 1 or 0, corresponding to the singlet and triplet projections, respectively. In particular, the singlet reservoir measures the observable ${\rm Q_S}$ at the rate $k_{\rm S}/2$. The "yes" result of this measurement corresponds to ${\rm Q_S}=1$ and the singlet projection, while the no/null result corresponds to the triplet projection. Similarly, the triplet reservoir measures the observable ${\rm Q_T}=\mathbbmtt{1}-{\rm Q_S}$ at the rate $k_{\rm T}/2$. The "yes" result of this measurement corresponds to ${\rm Q_S}=0$ and a triplet projection, while the no/null result corresponds to the singlet projection. Equivalently, ${\rm Q_S}$ is measured at the total rate $(k_{\rm S}+k_{\rm T})/2$. Again, these measurements are unobservable and lead to the aforementioned S-T dephasing. What is observable is the detection of a neutral recombination product. The corresponding null detection implies the possibilities (i)-(iii).

For testing our code and providing a "baseline" for the simulations of Section VII we show in Fig.\ref{fig3} an example of an MC simulation of just the singlet-triplet decoherence described by \eqref{MEnr}. 
To simulate the quantum trajectories of non-recombining RPs we start with $10^4$ RPs all being in the singlet state at $t=0$. We then evolve the state of each RP, using in each time increment $dt$ a random number $r$ uniformly distributed between 0 and 1. If $r<dp_{\rm S}$ we project the RP trajectory to the singlet state, if $dp_{\rm S}<r<dp_{\rm S}+dp_{\rm T}$ we project it to the triplet state, and if $1>r>dp_{\rm S}+dp_{\rm T}$ we evolve the RP state with the Hamiltonian ${\cal H}$. Due to these random quantum jumps, the S-T oscillations of the RPs suffer dephasing, hence the trajectory-averaged  expectation value of ${\rm Q_S}$ exhibits S-T oscillations of decaying amplitude. The perfect agreement between MC and the master equation \eqref{MEnr} shown in Fig.\ref{fig3}b is expected {\it by definition}, i.e. the physics included in the MC simulation are those exactly reproducing the master equation. This agreement does not convey any information other than that our code is working properly and that the 10000 trajectories are statistically adequate for the comparison undertaken in the following.
\section{Singlet-Triplet Coherence}
Since ${\rm Q_S}+{\rm Q_T}=\mathbbmtt{1}$ (the unit matrix is henceforth understood to have the dimension of the particular RP under consideration), any density matrix $\rho$ can be written as $\rho=({\rm Q_S}+{\rm Q_T})\rho({\rm Q_S}+{\rm Q_T})$, or 
\beq
\rho=\rho_{\rm SS}+\rho_{\rm TT}+\rho_{\rm ST}+\rho_{\rm TS},\label{generalrho}
\eeq
where $\rho_{xy}=Q_{x}\rho Q_{y}$, with $x,y={\rm S,T}$. It is clear that $\rho_{\rm SS}+\rho_{\rm TT}$ forms the incoherent  part of $\rho$, whereas the S-T coherence is represented by $\rho_{\rm ST}+\rho_{\rm TS}$. A naturally arising question is how coherent is a particular RP state described by some density matrix $\rho$. Consider for simplicity an imaginary 4-dimensional RP. The state $|\psi\rangle=(|{\rm S}\rangle+|{\rm T}_{0}\rangle)/\sqrt{2}$, or equivalently $\rho={1\over 2}|{\rm S}\rangle\langle {\rm S}|+{1\over 2}|{\rm T}_0\rangle\langle {\rm T}_0|+{1\over 2}|{\rm S}\rangle\langle {\rm T}_{0}|+{1\over 2}|{\rm T}_{0}\rangle\langle {\rm S}|$ clearly is maximally S-T coherent, whereas the state $\rho={1\over 2}|{\rm S}\rangle\langle {\rm S}|+{1\over 2}|{\rm T}_0\rangle\langle {\rm T}_0|$ is maximally incoherent. There could also be an intermediate  case of partial coherence, such as $\rho={1\over 2}|{\rm S}\rangle\langle {\rm S}|+{1\over 2}|{\rm T}_0\rangle\langle {\rm T}_0|+a|{\rm S}\rangle\langle {\rm T}_{0}|+a|{\rm T}_{0}\rangle\langle {\rm S}|$, with $a<1/2$. We thus need a measure of the "strength" of the "off-diagonal part" $\rho_{\rm ST}$ of the density matrix. In \cite{komPRE2011} we introduced the measure of coherence 
\beq
p_{\rm coh}(\rho)={{\tr\{\rho_{\rm ST}\rho_{\rm TS}\}}\over {\tr\{\rho_{\rm SS}\}\tr\{\rho_{\rm TT}\}}}\label{pcoh}
\eeq
However, this definition of $p_{\rm coh}$ is flawed in the following sense. S-T coherence is reflected by the value of the off-diagonal elements of the density matrix in the S-T basis. It is intuitively expected that such a measure should scale
linearly with the off-diagonal elements, however $p_{\rm coh}$ scales as the square of the off-diagonal elements of $\rho$. Hence if they decay at some rate $\Gamma$, $p_{\rm coh}$ will decay at $2\Gamma$, and this will skew the description of the relevant dynamics. 
\subsection{Rigorous analysis of S-T coherence}
Although essential, a rigorous quantification of coherence in quantum systems has received little attention, at least compared to the quantification of entanglement which has advanced through the definition of several measures \cite{entmeas1,entmeas2}. Recently, Plenio and co-workers introduced a rigorous approach to quantifying quantum coherence \cite{plenio_coherence}. We will follow this approach to introduce a new well-behaved measured of S-T coherence.

The first step is to define the set of incoherent states $\mathcal{I}$. Since we are interested in S-T coherence, it is straightforward to define $\mathcal{I}$ as the set containing all density matrices $\rho$ for which $\rho=\rho_{\rm SS}+\rho_{\rm TT}$, i.e. the coherences $\rho_{\rm ST}$ and $\rho_{\rm TS}$ are absent. Plenio and coworkers then define a set of three criteria that any measure of coherence should satisfy. The first and most obvious (and the one that will be used in the following) is that $p_{\rm coh}(\rho)=0$ for $\rho\in\mathcal{I}$. In order not to overburden this discussion with technical details, this and the other two criteria are reproduced in Appendix A, where we also demonstrate in more detail the shortcomings of our previous definition \eqref{pcoh}. 

In the new definition of $p_{\rm coh}$ to be shortly introduced, $p_{\rm coh}$  scales linearly with the off-diagonal elements of $\rho$, as it conforms with the Hilbert-Schmidt norm $C_{l_1}(\rho)$ shown in \cite{plenio_coherence} to be an acceptable measure of coherence.  In this measure Plenio and co-workers sum the absolute value of all off-diagonal elements of the density matrix. However, we are not interested in quantifying coherences within the triplet subspace, e.g. among $|{\rm T}_{+}\rangle$ and $|{\rm T}_{-}\rangle$. Neither are we interested in nuclear spin coherences. We are only concerned with the coherence between the electron singlet and triplet subspaces. So in our new definition we will sum the absolute value of the amplitudes appearing in the coherences $|S\rangle\langle {\rm T}_{0}|$, $|{\rm S}\rangle\langle {\rm T}_{+}|$ and $|{\rm S}\rangle\langle {\rm T}_{-}|$. To do so we define
\beq
{\cal C}(\rho)=\sum_{j=0,\pm}\sqrt{\tr\{\rho_{\rm ST}|T_j\rangle\langle T_j|\rho_{\rm TS}\}}\label{crho}
\eeq
This definition is visualized by a simple example in Appendix B. Before defining the new measure $p_{\rm coh}$ we note the following: (i) since $\tr\{\rho\}$ is a decaying function of time due to recombination, we have to normalize ${\cal C}(\rho)$ by $\tr\{\rho\}$ in order to get the genuine measure of coherence for the surviving RPs. (ii) as mentioned in  \cite{plenio_coherence} the state of maximum coherence in a $d$-dimensional Hilbert space with basis $|j\rangle$ is $\sum_{j=1}^{d}{1\over \sqrt{d}}|j\rangle$. In our case, the most general pure state of an RP can be written as $|\psi\rangle=\alpha_{\rm S}|{\rm S}\rangle\otimes|\chi_{\rm S}\rangle+\sum_{j=0,\pm}\alpha_{j}|{\rm T}_{j}\rangle|\chi_{j}\rangle$, where $|\chi_{\rm S}\rangle$ and $|\chi_{j}\rangle$ are normalized nuclear spin states. Here S-T coherence is maximum when $|\alpha_{\rm S}|=|\alpha_{j}|=1/2$, and this maximum value is $\sum_{j=0,\pm}|\alpha_{\rm S}\alpha_{j}|=3/4$. However, if the Hamiltonian excites a subset of these coherences, e.g. only the S-T$_0$ coherence, the maximum value of the coherence would be smaller. Since in the following we use $p_{\rm coh}$ as a probability measure, we normalized ${\cal C}(\rho)$ with its maximum value obtained when $\rho$ evolves unitarily under the action of ${\cal H}$. So now we define 
\beq
p_{\rm coh}(\rho)={1\over {\tr\{\rho\}}}{{{\cal C}(\rho)}\over {\rm max\{{\cal C}(\tilde{\rho})\}}}\label{pcohnew}
\eeq
where $d\tilde{\rho}/dt=-i[{\cal H},\tilde{\rho}]$. We note that this new definition of $p_{\rm coh}$ is numerically very similar to the square-root of our earlier definition \eqref{pcoh}.
\section{Definition of $\rho_{\rm coh}$ and $\rho_{\rm incoh}$}
It is clear from \eqref{crho} that if we scale $\rho_{\rm ST}$ and $\rho_{\rm TS}$ with a positive number $\lambda$, i.e. if $\rho_{\rm ST}\rightarrow\lambda\rho_{\rm ST}$ and $\rho_{\rm TS}\rightarrow\lambda\rho_{\rm TS}$ then $p_{\rm coh}\rightarrow\lambda p_{\rm coh}$. 
So going back to the general form \eqref{generalrho} of the density matrix $\rho$, if we choose $\lambda=1/p_{\rm coh}$, that is, if we define the density matrix
\beq
\rho_{\rm coh}=\rho_{\rm SS}+\rho_{\rm TT}+{1\over p_{\rm coh}}\rho_{\rm ST}+{1\over p_{\rm coh}}\rho_{\rm TS}\label{rhocoh},
\eeq
then $\rho_{\rm coh}$ will describe a maximally coherent state, $p_{\rm coh}(\rho_{\rm coh})=1$. The density matrix $\rho_{\rm coh}$ can be thought of as the S-T coherence distillation of $\rho$. 
We can also define a maximally incoherent density matrix $\rho_{\rm incoh}$: 
\beq
\rho_{\rm incoh}=\rho_{\rm SS}+\rho_{\rm TT},\label{rhoincoh}
\eeq
for which $p_{\rm coh}(\rho_{\rm incoh})=0$. Using Eqs. \eqref{generalrho}, \eqref{rhocoh} and \eqref{rhoincoh} it is then trivial to show that {\it any} density matrix $\rho$ can be written as:
\beq
\rho=(1-p_{\rm coh})\rho_{\rm incoh}+p_{\rm coh}\rho_{\rm coh}\label{decomp}
\eeq
This will be the starting point for the retrodictive derivation presented in the following Section. We note that this general decomposition of $\rho$ into $\rho_{\rm incoh}$ and $\rho_{\rm coh}$ was possible due to the particular definition of $\rho_{\rm coh}$ and its property that $p_{\rm coh}(\rho_{\rm coh})=1$, which itself relies on the linear scaling of $p_{\rm coh}$ mentioned previously. In other words, the following formal derivation based on quantum retrodiction would not be possible without the proper definition of the S-T coherence measure.
\section{Quantum retrodiction and radical-ion-pair recombination}
\subsection{Radical-ion-pair recombination from the single-molecule and from the ensemble perspective}
The density matrix of an ensemble of $N$ RPs is $\rho_t=\sum_{i=1}^{N}|\psi_i(t)\rangle\langle\psi_i(t)|$, where $|\psi_{i}\rangle$ is the spin state of the $i$-th RP. Each $|\psi_i\rangle$ has suffered a number of S- or T-quantum jumps until the time $t$. Due to recombination $N$ is time-dependent, since if the $i$-th RP recombines at time $t$, its quantum state $|\psi_i\rangle\langle\psi_{i}|$ at time $t$ must be subtracted from $\rho_t$ in order to update $\rho_t$ into $\rho_{t+dt}$. Although this is a simple physical picture from the perspective of quantum trajectories, it is not straightforward to translate it into a master equation. The root of the difficulty is S-T dephasing, which transforms a pure initial state into a mixture. 

As well known, there is no unique way to unravel a density matrix into its component pure states. Hence we have to make due with the following physical scenario. Given the density matrix $\rho_t$ at some time $t$, and given the {\it measured} singlet and triplet neutral products during the infinitesimal interval $dt$, $dn_{\rm S}$ and $dn_{\rm T}$, respectively, how do we update $\rho_t$ into $\rho_{t+dt}$? In general, the change $d\rho=\rho_{t+dt}-\rho_t$ is caused by (i) the change of state of RPs that did not recombine during $dt$, call it $d\rho_{\rm decoh}$, given by \eqref{MEnr} and (ii) the RPs that did recombine during $dt$, call it $d\rho_{\rm recomb}$, i.e. $d\rho=d\rho_{\rm decoh}+d\rho_{\rm recomb}$. Clearly, $\tr\{d\rho\}=\tr\{d\rho_{\rm recomb}\}=-dn_{\rm S}-dn_{\rm T}$, but that alone cannot lead to the form of $d\rho_{\rm recomb}$. 

We will now derive $d\rho_{\rm recomb}$ using the formal tools of quantum retrodiction. We then compare the predictions of the new master equation to the Monte Carlo simulation. The latter turns out to be a very useful tool, since dealing with an ensemble of pure states allows us, by default, to subtract the particular component states $|\psi_i\rangle$ of the recombined RPs from the considered ensemble.
\subsection{Radical-ion-pair recombination and quantum retrodiction}
The predictive approach to quantum measurements, which we are most familiar with, addresses the question: given the density matrix describing a physical system, what are the probabilities of specific measurement outcomes? The so-called retrodictive approach \cite{retro1,retro2}, used less often, is about the reverse: given a specific measurement outcome, what is the probability that the system's state prior to the measurement was this or that? Quantum retrodiction is relevant to quantum communication \cite{retro3,retro4}, since Bob, the receiver of quantum information, attempts to reconstruct the quantum state delivered to him by Alice, the sender, based on specific measurement outcomes. 

The idea relating RP recombination to the concept of retrodiction and S-T coherence is the following. When an RP is in a particular state $|\psi\rangle$ just before it recombines, we must subtract $|\psi\rangle\langle\psi|$ from the density matrix to account for this recombination event. But since S-T dephasing produces a mixture of pure states, given the recombination product, which is either the singlet or the triplet ground state, one cannot unambiguously retrodict the pre-recombination state $|\psi\rangle$. A singlet recombination could for example result from a singlet RP as much as from an S-T coherent RP. The theory of quantum retrodiction allows us to retrodict $|\psi\rangle$ "on average". The way this is done depends on how coherent is the RP state described by the density matrix $\rho$, hence the necessity of defining $p_{\rm coh}$. 

This is seen by examining the two extreme cases of minimum and maximum S-T coherence, for which $d\rho_{\rm recomb}$ is straightforward to derive.
Based on the general decomposition \eqref{decomp}, the theory of quantum retrodiction can then be seamlessly applied in the general case of a density matrix with partial S-T coherence.
\subsection{Recombination of maximally coherent radical-ion pairs}
Suppose that at time $t$ we have an ensemble of $N$  RPs all in some maximally S-T coherent state $|\psi\rangle$. Suppose further that the only change during the interval $dt$ is the recombination of just one RP, i.e. the detection of one neutral product. Clearly, scaling the normalization of $\rho$ from 1 to $N$ just for the sake of this discussion, it is $\rho_t=N|\psi\rangle\langle\psi |$ and $\rho_{t+dt}=(N-1)|\psi\rangle\langle\psi |$, since now we have one less RP in the state $|\psi\rangle$. This can be formalized as follows. For a maximally coherent ensemble of RPs all in the same state $|\psi\rangle$, the single-molecule density matrix will be $\rho/\tr\{\rho\}$. If we define $\delta\rho_{\rm coh}^{\rm 1S}$ ($\delta\rho_{\rm coh}^{\rm 1T}$) to be the change in $\rho$ due to the measurement of {\it just one} singlet (triplet) neutral product, it will be
\beq
\delta\rho_{\rm coh}^{\rm 1S}=\delta\rho_{\rm coh}^{\rm 1T}=-{\rho\over {\tr\{\rho\}}}\label{drcoh}
\eeq 
\subsection{Recombination of maximally incoherent radical-ion pairs}
In the other extreme, suppose that $\rho_t$ is a maximally incoherent mixture of singlet and triplet RPs, i.e. $\rho_t=\rho_{\rm SS}+\rho_{TT}$. Then the detection of a singlet (triplet) recombination product leads us to conclude with certainty that it resulted from a singlet (trilpet) RP and hence we can reduce the population of singlet (triplet) RPs by one. If we define $\delta\rho_{\rm incoh}^{\rm 1S}$ ($\delta\rho_{\rm incoh}^{\rm 1T}$) to be the change in $\rho$ due to the recombination of {\it just one} singlet (triplet) RP, it will be
\begin{align}
\delta\rho_{\rm incoh}^{\rm 1S}&=-{{{\rm Q_S}\rho{\rm Q_S}}\over {\tr\{ {\rm Q_S}\rho {\rm Q_S}\}}}=-{{{\rm Q_S}\rho{\rm Q_S}}\over {\tr\{\rho {\rm Q_S}\}}}\label{drincohS}\\
\delta\rho_{\rm incoh}^{\rm 1T}&=-{{{\rm Q_T}\rho{\rm Q_T}}\over {\tr\{ {\rm Q_T}\rho {\rm Q_T}\}}}=-{{{\rm Q_T}\rho{\rm Q_T}}\over {\tr\{\rho {\rm Q_T}\}}}\label{drincohT}
\end{align}
The last equality in the above equations follows from the cyclic property of the trace and the fact that ${\rm Q_S}$ and ${\rm Q_T}$ are projectors, hence idempotent.
\subsection{Recombination of radical-ion pairs having partial S-T coherence}
We will now use the formalism of quantum retrodiction to derive the reaction operators for the general case of partial S-T coherence. The retrodiction formalism \cite{retro3,retro4} uses the preparation operators $\Lambda_{i}$ and the measurement operators $\Pi_{j}$. In particular, suppose that a system is prepared in a state $\rho_{i}$ with probability $P(i)$.  The preparation operator is then defined as $\Lambda_{i}=P(i)\rho_{i}$. If the particular preparation is unknown then we have to average over all possible preparations and the system will be described by the density matrix $\rho=\sum_{i}\Lambda_{i}$. Suppose further that a measurement defined by the POVM set $\Pi_{i}$, where $\sum_{i}\Pi_{i}=\mathbbmtt{1}$, returns the $j$-th result. Defining $\rho_{j}^{r}=\Pi_{j}/\tr\{\Pi_{j}\}$, the main result of retrodiction theory is that the {\it conditional probability} that state $\rho_{i}$ was prepared, given the measurement result $j$ is 
\beq
P(i|j)={{\tr\{\Lambda_{i}\rho_{j}^{r}\}}\over {\sum_{i}\tr\{\Lambda_{i}\rho_{j}^{r}\}}}\label{retr}
\eeq
The POVM set of measurement operators of interest in our case consists of $\Pi_{1}={\rm {\rm Q_S}}$ and $\Pi_{2}={\rm Q_T}$, already mentioned to satisfy the condition ${\rm {\rm Q_S}}+{\rm Q_T}=\mathbbmtt{1}$. As shown before, the general form of the RP density matrix at time $t$ can be written as $\rho=\Lambda_{1}+\Lambda_{2}=(1-p_{\rm coh})\rho_{\rm incoh}+p_{\rm coh}\rho_{\rm coh}$, i.e. we identify $\Lambda_{1}=(1-p_{\rm coh})\rho_{\rm incoh}$ and $\Lambda_{2}=p_{\rm coh}\rho_{\rm coh}$, where $\rho_{\rm coh}$ and $\rho_{\rm incoh}$ have been defined by \eqref{rhocoh} and \eqref{rhoincoh}, respectively. 

Suppose that during the interval $dt$ we have detected one $x$ neutral product, where $x={\rm S,T}$. To apply Eq. \eqref{retr}, we note that since $\rho_{x}^{r}={\rm Q}_x/\tr\{{\rm Q}_x\}$, the denominator $\tr\{{\rm {\rm Q}_x}\}$ of $\rho_{x}^{r}$ will drop out of Eq. \eqref{retr}. Further, since $\rho=\sum_{i}\Lambda_{i}$, the denominator in Eq. \eqref{retr} is proportional to the expectation value of ${\rm Q}_x$ at time $t$, i.e. $\sum_{i}\tr\{\Lambda_{i}\rho_{x}^{r}\}\propto\tr\{\rho {\rm {\rm Q}_x}\}$, hence given the detection of one $x$ neutral product, the probabilities that it originated either from $\rho_{\rm incoh}$ or from $\rho_{\rm coh}$ are
\begin{align}
P({\rm incoh}|x)&={{\tr\{\Lambda_{1}{\rm {\rm Q}_x}\}}\over {\tr\{\rho{\rm {\rm Q}_x}\}}}=(1-p_{\rm coh}){{\tr\{\rho_{\rm incoh} {\rm {\rm Q}_x}\}}\over {\tr\{\rho{\rm Q}_x\}}}\nonumber\\
P({\rm coh}|x)&={{\tr\{\Lambda_{2}{\rm {\rm Q}_x}\}}\over {\tr\{\rho{\rm {\rm Q}_x}\}}}=p_{\rm coh}{{\tr\{\rho_{\rm coh} {\rm {\rm Q}_x}\}}\over {\tr\{\rho{\rm Q}_x\}}} 
\end{align}
Since the expectation value of ${\rm Q}_x$ in $\rho$ is the same as in $\rho_{\rm incoh}$ and $\rho_{\rm coh}$, it readily follows that 
\begin{align}
P({\rm incoh}|{\rm S})&=P({\rm incoh}|{\rm T})=1-p_{\rm coh}\nonumber\\
P({\rm coh}|{\rm S})&=P({\rm coh}|{\rm T})=p_{\rm coh}\nonumber
\end{align}
We have shown how the density matrix changes upon detecting just one product in the extreme cases of maximum/minimum coherence. In the general case when the RP ensemble is described by $\rho$, detecting 
{\it just one} singlet (triplet) neutral product leads to a change in $\rho$ given by $\delta\rho^{\rm 1S}$ ($\delta\rho^{\rm 1T}$), where
\begin{align}
\delta\rho^{\rm 1S}&=P({\rm incoh}|{\rm S})\delta\rho_{\rm incoh}^{\rm 1S}+P({\rm coh}|{\rm S})\delta\rho_{\rm coh}^{\rm 1S}\nonumber\\
\delta\rho^{\rm 1T}&=P({\rm incoh}|{\rm T})\delta\rho_{\rm incoh}^{\rm 1T}+P({\rm coh}|{\rm T})\delta\rho_{\rm coh}^{\rm 1T}\nonumber
\end{align}
The generalization to the case of detecting $dn_{\rm S}=k_{\rm S}dt\tr\{\rho{\rm Q_S}\}$ singlet and $dn_{\rm T}=k_{\rm T}dt\tr\{\rho{\rm Q_T}\}$ triplet neutral products is now straightforward:
\begin{align}
d\rho_{\rm recomb}&=dn_{\rm S}\delta\rho^{\rm 1S}+dn_{\rm T}\delta\rho^{\rm 1T}\label{recomb}
\end{align}
Since $\tr\{\delta\rho_{\rm coh}^{\rm 1S}\}=\tr\{\delta\rho_{\rm coh}^{\rm 1T}\}=\tr\{\delta\rho_{\rm incoh}^{\rm 1S}\}=\tr\{\delta\rho_{\rm incoh}^{\rm 1T}\}=-1$, it is $\tr\{d\rho_{\rm recomb}\}=-dn_{\rm S}-dn_{\rm T}$, as it should be. 

Using \eqref{MEnr} and \eqref{recomb}, we arrive at the master equation describing RP quantum dynamics:
\begin{align}
{{d\rho}\over {dt}}=&-i[{\cal H},\rho]\label{t1}\\  
&-{{k_{\rm S}+k_{\rm T}}\over 2}\big(\rho {\rm {\rm Q_S}}+{\rm {\rm Q_S}}\rho-2{\rm {\rm Q_S}}\rho {\rm {\rm Q_S}}\big)\label{t2}\\
&-(1-p_{\rm coh})\big(k_{\rm S}{\rm {\rm Q_S}}\rho {\rm {\rm Q_S}}+k_{\rm T} {\rm Q_T} \rho {\rm Q_T}\big)\label{t3}\\
&-p_{\rm coh}{{dn_{\rm S}+dn_{\rm T}}\over {dt}}{\rho_{\rm coh}\over {\tr\{\rho\}}}\label{t4}
\end{align}
The term in \eqref{t1} is the unitary Hamiltonian evolution which generates S-T coherence, the dissipation of which is given by term \eqref{t2}, while \eqref{t3} and \eqref{t4} are the spin-dependent reaction terms. This master equation has a form identical to the one derived in \cite{komPRE2011}, the crucial difference being the new definition of $p_{\rm coh}$ and the last term \eqref{t4} where we now have the appearance of $\rho_{\rm coh}$ instead of $\rho$ that was used phenomenologically in \cite{komPRE2011}. 

Finally, we rewrite the master equation \eqref{t1}-\eqref{t4} in a more "user-friendly" form involving only the matrices $\rho_{xy}={\rm Q}_{x}\rho {\rm Q}_{y}$, where $x,y={\rm S,T}$.
\begin{align}
{{d\rho}\over {dt}}=&-i[{\cal H},\rho]\nonumber\\  
&-{{k_{\rm S}+k_{\rm T}}\over 2}\big(\rho_{\rm ST}+\rho_{\rm TS}\big)\nonumber\\
&-(1-p_{\rm coh})\big(k_{\rm S}\rho_{\rm SS}+k_{\rm T}\rho_{\rm TT}\big)\nonumber\\
&-{1\over {\tr\{\rho\}}}\big(k_{\rm S}\tr\{\rho_{\rm SS}\}+k_{\rm T}\tr\{\rho_{\rm TT}\}\big)\times\nonumber\\
&~~~\big(p_{\rm coh}\rho_{\rm SS}+p_{\rm coh}\rho_{\rm TT}+\rho_{\rm ST}+\rho_{\rm TS}\big)\nonumber
\end{align}
\begin{figure}
\includegraphics[width=8.0 cm]{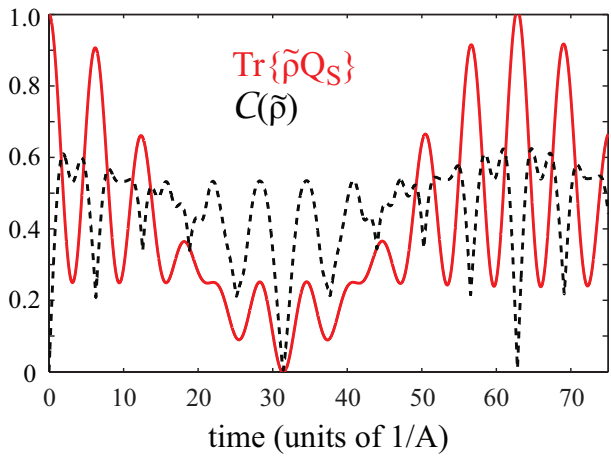}
\caption{(Color online) Time evolution of ${\tr\{\tilde{\rho}\rm Q_S}\}$ (red solid line) and S-T coherence ${\cal C}(\tilde{\rho})$ (black dashed line) for the same RP considered in Fig. 3, taking into account only S-T mixing driven by the Hamiltonian ${\cal H}$, i.e. $d\tilde{\rho}/dt=-i[{\cal H},\tilde{\rho}]$. The singlet state obviously corresponds to zero S-T coherence, while the state in-between the extrema of ${\tr\{\tilde{\rho}\rm Q_S}\}$ corresponds to an S-T superposition and hence maximum S-T coherence.}
\label{fig4}
\end{figure}
\begin{figure}
\includegraphics[width=8.5 cm]{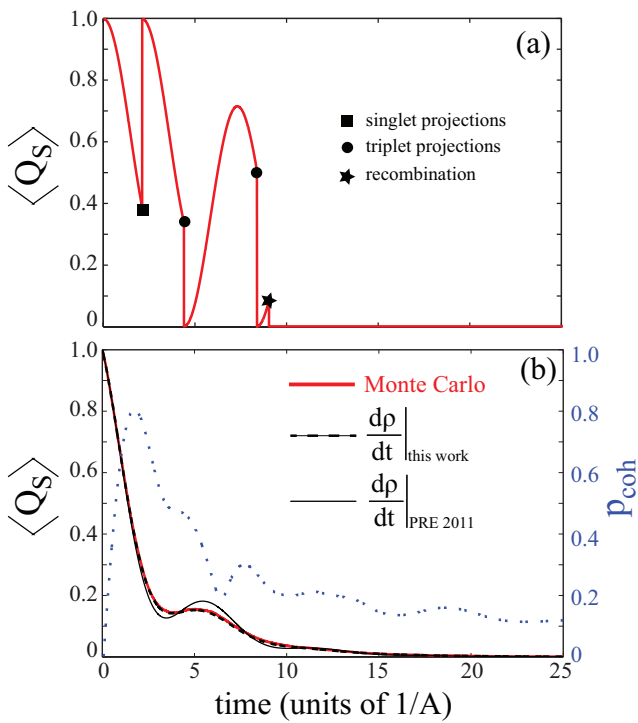}
\caption{(Color online) Time evolution of $\langle{\rm Q_S}\rangle$ including S-T mixing, S-T decoherence and recombination for the same RP Hamiltonian used in Figs.\ref{fig3}-\ref{fig4}, with $k_{\rm S}=k_{\rm T}=A/4$. (a) example of a single-RP quantum trajectory with initial state $|S\rangle\otimes|\uparrow\rangle$. (b) Monte Carlo simulation (red solid line) using 10,000 trajectories (two initial states $|S\rangle\otimes|\uparrow\rangle$ and $|S\rangle\otimes|\downarrow\rangle$, with 5000 trajectories for each), prediction of the master equation of this work (dashed line), and the earlier theory (solid line) introduced in \cite{komPRE2011}. The corresponding measure of S-T coherence $p_{\rm coh}$ is shown with the blue dotted line. The Monte Carlo and the theoretical prediction of this work coincide.}
\label{fig5}
\end{figure}
\section{Monte Carlo simulation of S-T decoherence and recombination using single-molecule quantum trajectories}
To the simulation presented in Section III we now add two additional possibilities in each time step $dt$: singlet and triplet recombination with probability $dr_{\rm S}=k_{\rm S}dt\langle{\rm Q_S}\rangle$ and $dr_{\rm T}=k_{\rm T}dt\langle{\rm Q_T}\rangle$, respectively. In the event that the $j$-th RP recombines within $dt$ at time $t$, its state $|\psi_{j}\rangle\langle\psi_{j}|$ is subtracted at time $t$ from the sum $\rho=\sum_{i}|\psi_{i}\rangle\langle\psi_{i}|$. 

For a more comprehensive understanding of the considerations to follow, we first show in Fig.\ref{fig4} just the Hamiltonian evolution (no decoherence, no reaction) of $\langle{\rm Q_S}\rangle=\tr\{\tilde{\rho}{\rm Q_S}\}$ and ${\cal C}(\tilde{\rho})$ for the model RP considered in our numerical examples. Clearly, when $\langle{\rm Q_S}\rangle=1$ it is ${\cal C}(\tilde{\rho})=0$, as expected since we have no coherence between the singlet and triplet subspace. This coherence is maximum at intermediate times in-between the extrema of $\langle {\rm Q_S}\rangle$. 

In Fig.\ref{fig5}a we depict a single-RP quantum trajectory, similar to the one shown in Fig.\ref{fig3}a but now also including recombination. The recombination rates are taken equal, $k_{\rm S}=k_{\rm T}$. In Fig.\ref{fig5}b we show that using the newly derived master equation \eqref{t1}-\eqref{t4} we obtain a perfect agreement with the MC simulation that was lacking with the earlier theory. The MC simulation is the average of $10^4$ trajectories like the one shown in Fig.\ref{fig5}a. In Fig.\ref{fig5}b we also include the time evolution of $p_{\rm coh}$.

\begin{figure}
\includegraphics[width=8.5 cm]{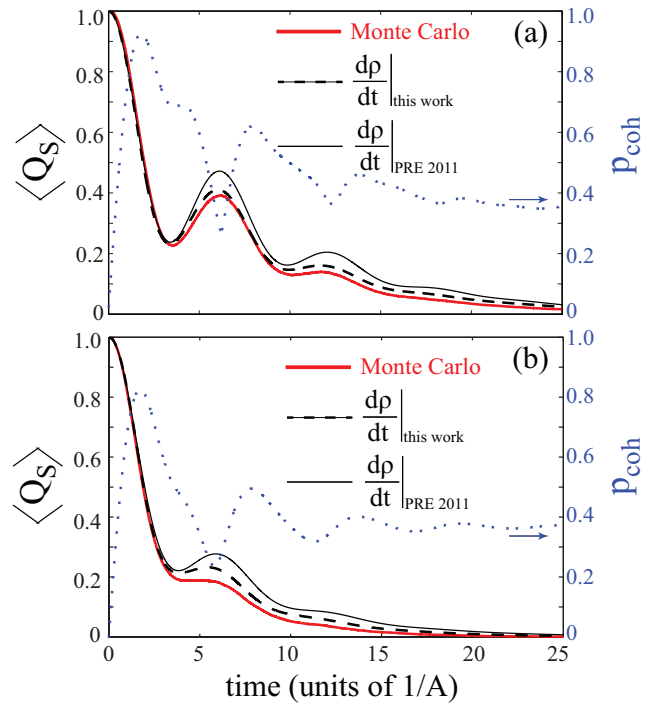}
\caption{(Color online) Similar plots with Fig.\ref{fig5} but with asymmetric recombination rates. (a) $k_{\rm S}=0$, $k_{\rm T}=A/4$. (b) $k_{\rm S}=0$, $k_{\rm T}=A/2$.}
\label{fig6}
\end{figure}

We next move to the asymmetric regime where $k_{\rm T}\neq 0$ and $k_{\rm S}=0$. This is of interest as it is found in the RPs appearing in a large number of photosynthetic reaction centers \cite{matysik}. In Fig.\ref{fig6}a and Fig.\ref{fig6}b we again plot $\langle{\rm Q_S}\rangle$ for $k_{\rm S}=0$, $k_{\rm T}=A/4$ and $k_{\rm T}=A/2$, respectively. While for the former we get a very good agreement between the Monte Carlo simulation and the master equation, the agreement is not perfect for the latter, but still much better than our earlier theory.
We comment on this in Section IX.
\section{Decay rate of singlet-triplet coherence}
\begin{figure}
\includegraphics[width=8.5 cm]{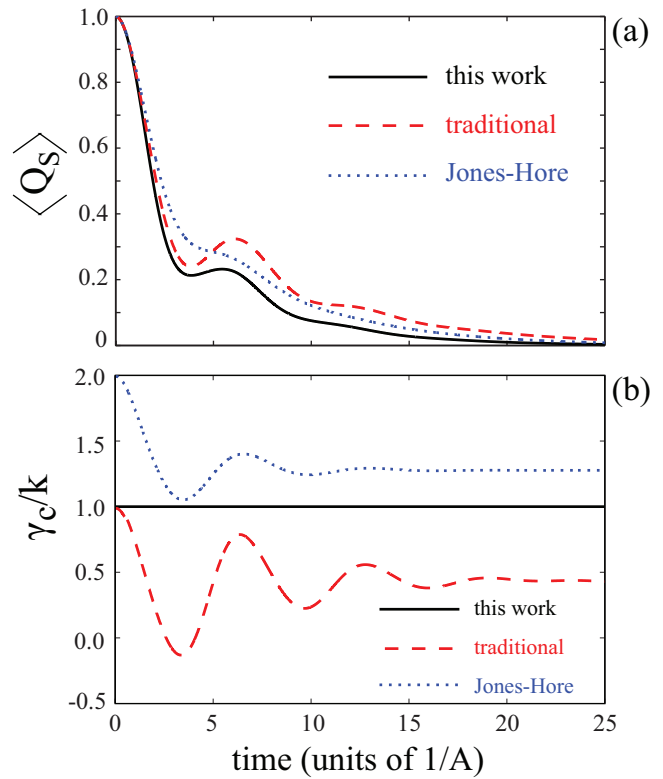}
\caption{(Color online) Comparison of the three theories for the case presented in Fig.\ref{fig6}b, i.e. $k_{\rm S}=0$ and $k_{\rm T}=A/2$. (a) the S-T coherence, embodied by the amplitude of the oscillation of $\langle{\rm Q_S}\rangle$ decays faster in the Jones-Hore theory, slower in our theory and even slower in the traditional theory. (b) the corresponding decay rate $\gamma_{c}/k$, where $k_{\rm T}=2k$.}
\label{fig7}
\end{figure}

For the sake of completeness we present a comparison between our theory, the traditional (or Haberkorn) approach \cite{haberkorn} and the theory put forward by Jones \& Hore \cite{JH}. First we reiterate \cite{komPRE2011} that the traditional theory results from our theory by forcing $p_{\rm coh}=0$. We also note that our master equation \eqref{t1}-\eqref{t4} is identical with the Jones-Hore equation in the case $k_{\rm S}=k_{\rm T}$. In this special case $p_{\rm coh}$ drops out of our master equation \eqref{t1}-\eqref{t4}. In Fig.\ref{fig7}a we plot the time evolution of $\langle{\rm Q_S}\rangle$ for all three theories, which qualitatively look quite similar. Their most obvious difference is how fast the S-T coherence is lost. By inspection it readily appears that the amplitude of the S-T oscillations in Fig.\ref{fig7}a decays faster in the Jones-Hore theory, slower in our theory and even slower in the traditional approach. We will now rigorously quantify this observation by following a general approach equally applicable to all three theories. This is based on the general decomposition \eqref{generalrho}, in particular we will consider the coherent part of $\rho$ which is $\rho_{c}=\rho_{\rm ST}+\rho_{\rm TS}$.
In our master equation $\rho_{c}$ appears both in the term \eqref{t2} and in the term \eqref{t4}. The latter is obvious, while the former can be seen by simple operator manipulations leading to ${\rm Q_S}\rho+\rho{\rm Q_S}-2{\rm Q_S}\rho{\rm Q_S}=\rho_{c}$. Thus, if we right (left) multiply the master equation \eqref{t1}-\eqref{t4} with ${\rm Q_S}$ (${\rm Q_T}$), then vice-versa, and take the sum we find that $\rho_{c}$ obeys the equation
\beq
{{d\rho_{c}}\over{dt}}=-i[{\cal H},\rho]_{c}-\Gamma_{c}\rho_{c}\label{rc},
\eeq
where $[{\cal H},\rho]_{c}={\rm Q_S}[{\cal H},\rho]{\rm Q_T}+{\rm Q_T}[{\cal H},\rho]{\rm Q_S}$. The decay of $\rho_{c}$ is governed by the rate 
\beq
\Gamma_{c}=k_{\rm S}\Big({1\over 2}+\langle{\rm \tilde{Q}_S}\rangle\Big)+k_{\rm T}\Big({1\over 2}+\langle{\rm \tilde{Q}_T}\rangle\Big),
\eeq
where we defined $\langle{\rm \tilde{Q}_x}\rangle=\tr\{\rho{\rm Q}_x\}/\tr\{\rho\}$ with $x={\rm S,T}$. Moreover, since it will be needed in the following, by taking the trace of both sides in \eqref{t1}-\eqref{t4} we find that $\tr\{\rho\}$, the normalization of $\rho$, obeys the equation 
\beq
{{d\tr\{\rho\}}\over{dt}}=-\kappa\tr\{\rho\}\label{trr}, 
\eeq
where
\beq
\kappa=k_{\rm S}\langle{\rm \tilde{Q}_S}\rangle+k_{\rm T}\langle{\rm \tilde{Q}_T}\rangle\label{kappa}
\eeq
We finally define the "genuine" S-T decoherence rate as $\gamma_{c}=\Gamma_{c}-\kappa$. This describes the decay of S-T coherence due to all effects other than the changing normalization of $\rho$. This definition follows if we normalize $\rho_{c}$ by $\tr\{\rho\}$ and then use \eqref{rc} and \eqref{trr}.  Then we indeed find that the decay rate of $\rho_{c}/\tr\{\rho\}$ is $\gamma_c$.

We now consider two cases, (a) $k_{\rm S}=k_{\rm T}=k$, and (b) $k_{\rm S}=0$ and $k_{\rm T}=2k$, so that $k_{\rm S}+k_{\rm T}$ is the same in both cases. 
In case (a) we find that $\Gamma_{c}=2k$ since $\langle{\rm \tilde{Q}_S}\rangle+\langle{\rm \tilde{Q}_T}\rangle=1$. Moreover, $\kappa=k$, hence $\gamma_{c}=k$.
In case (b) it is $\Gamma_{c}=k(1+2\langle{\rm \tilde{Q}_T}\rangle)$, while $\kappa=2k\langle{\rm \tilde{Q}_T}\rangle$, hence $\gamma_{c}=k$.

We will now perform the same calculation for the traditional and the Jones-Hore theory. We first note that the equations \eqref{trr} and \eqref{kappa} are common for all three theories.
The traditional master equation is $d\rho/dt=-i[{\cal H},\rho]-k_{\rm S}({\rm Q_S}\rho+\rho{\rm Q_S})/2-k_{\rm T}({\rm Q_T}\rho+\rho{\rm Q_T})/2$. Again, multiplying from left and right with the projection operators as before we find that the decay rate of $\rho_{c}$ is $\Gamma_{c}=(k_{\rm S}+k_{\rm T})/2$. In case (a) it is found that $\gamma_{c}=0$, while in case (b) we get $\gamma_{c}=k(1-2\langle{\rm \tilde{Q}_T}\rangle)$. 
The Jones-Hore master equation is  $d\rho/dt=-i[{\cal H},\rho]-k_{\rm S}({\rm Q_S}\rho+\rho{\rm Q_S}-{\rm Q_S}\rho{\rm Q_S})-k_{\rm T}({\rm Q_T}\rho+\rho{\rm Q_T}-{\rm Q_T}\rho{\rm Q_T})$.
We similarly find that $\Gamma_{c}=k_{\rm S}+k_{\rm T}$. Then in case (a) it follows that $\gamma_{c}=k$ and in case (b) $\gamma_{c}=2k(1-\langle{\rm \tilde{Q}_T}\rangle)$. For clarity we summarize the results in Table I. 

The asymmetric case $k_{\rm T}\gg k_{\rm S}$ together with the singlet initial state is the regime of the quantum Zeno effect \cite{komPRE2009,pascazio,pascazio_review,kurizki} (most pronounced if $k_{\rm T}\gg\Omega$, where $\Omega$ is the S-T mixing frequency). In this regime, when the RP's spin state is about to evolve from the initial singlet state it is strongly back-projected to it due to the high $k_{\rm T}$. Thus, $\langle {\rm \tilde{Q}_S}\rangle$ decreases slowly from its initial value of 1, and hence $\langle {\rm \tilde{Q}_T}\rangle$ can be quite small, in particular, quite smaller than 1/2. This observation is common to all three theories. It thus follows that $2k(1-\langle\tilde{\rm Q}_{\rm T}\rangle)>k>k(1-2\langle\tilde{\rm Q}_{\rm T}\rangle)$. Indeed, as shown in Fig.\ref{fig7}b, the Jones-Hore theory predicts the largest decay rate for the S-T coherence, ours is intermediate and for the traditional theory it is the smallest.
\begin{table}
\caption{Decay rate of S-T coherence $\gamma_{c}$}
\begin{ruledtabular}
\begin{tabular}{|c|ccc|}
$\gamma_{c}$                               &   This work                                                  & Trad. theory                                                            & J.-H. theory\\
\hline
$k_{\rm S}=k_{\rm T}=k$              &                      $k$                                         &                                0                                                       &                   k                                           \\
$k_{\rm S}=0$, $k_{\rm T}=2k$   &                      $k$                                         &     $k(1-2\langle{\rm \tilde{Q}_T}\rangle)$                     & $2k(1-\langle{\rm \tilde{Q}_T}\rangle)$   \\
\end{tabular}
\end{ruledtabular}
\end{table}
\section{Discussion}
We  will finally comment on the success of the master equation \eqref{t1}-\eqref{t4} in matching the MC simulation, which has inbuilt the fundamental physical processes of RP reactions at the single-molecule level. While for the case $k_{\rm S}=k_{\rm T}$ there is a perfect agreement between theory and MC, independent of the particular definition of $p_{\rm coh}$, for the asymmetric case $k_{\rm T}\gg k_{\rm S}$ we have the more noticeable theory-MC a deviation the higher $k_{\rm T}$ is. For most practical purposes such a small deviation should be of little concern, however, it is worthwhile to discuss.

To our understanding, the problem is an underestimation of S-T coherence that in principle can be hardly overcome. The reason is the impossibility to unravel a density matrix into its component pure states. S-T decoherence will produce a mixture of S-T coherent, yet dephased states, which when described by a density matrix will look equivalent to a mixture of S-T incoherent and S-T coherent states, as we have shown with the decomposition into $\Lambda_1$ and $\Lambda_2$. To exacerbate the problem for the sake of this discussion, consider for example a mixture of the coherent states $|\psi_1\rangle=(|{\rm S}\rangle+|{\rm T}_0\rangle)/\sqrt{2}$ and $|\psi_2\rangle=(|{\rm S}\rangle-|{\rm T}_0\rangle)/\sqrt{2}$ with equal weights. Then $\rho={1\over 2}|\psi_1\rangle\langle\psi_1|+{1\over 2}|\psi_2\rangle\langle\psi_2|={1\over 2}(|{\rm S}\rangle\langle{\rm S}|+|{\rm T}_0\rangle\langle{\rm T}_0|)$. This state appears as maximally incoherent, yet it is formed by maximally coherent states. Having access to the information embodied by $\rho$, it is impossible to unravel or retrodict the constituents $|\psi_1\rangle$ or $|\psi_2\rangle$.

From \eqref{t3} it is seen that in the asymmetric case where $k_{\rm S}=0$, if $p_{\rm coh}$ is underestimated, then we remove a correspondingly larger triplet character from $\rho$, and hence $\rho$ appears to be more singlet than it really is, as is evident from Fig.\ref{fig6}, i.e. the master equation overshoots the MC. Moreover, this deviation is noticeable at the minima of $p_{\rm coh}$, while it is indiscernible at the maxima of $p_{\rm coh}$. Again, this is due to the reaction term \eqref{t3} of the master equation, which is more pronounced for low values of $p_{\rm coh}$.

We finally reiterate that what we have treated is the fundamental quantum dynamics of RP reactions governed by the physical processes inherent in the radical-pair mechanism, i.e. S-T dephasing and charge recombination, stemming from virtual and real transitions to the products' vibrational reservoirs, respectively. Clearly, other sources of decoherence could be present, which are either fundamental or technical, and the manifestation of which could depend on the physical realization of the RP dynamics, e.g. whether the molecules are in solution or in the solid state as in photosynthetic reaction centers. Dephasing due to a bath of surrounding nuclear spins that have not been included in the magnetic Hamiltonian has analogues in the study of quantum dots \cite{rosen,petta1,petta2} and has been considered by several authors \cite{kavokin,briegel,walters}. To our understanding, a consensus on the physical significance and the quantitative details of this hyperfine relaxation is still lacking from the literature. Whether the S-T dephasing we consider is a dominant process or not will at the end depend on the comparison between the particular recombination rates $k_{\rm S}$ and $k_{\rm T}$ of the RP under consideration and the hyperfine relaxation rate, or in general, the rates of other relaxation processes in the particular RP environment.

A detailed understanding of the interplay of all possible decoherence mechanisms, whether fundamental or technical, is outside the scope of this work. It is, however, a basic requirement for connecting the microscopic dynamics of RP reactions with behavioral observations of the avian compass mechanism, a non-trivial exercise recently undertaken in \cite{vedral,kaszli,gauger,kaszli2}.
 \section{Conclusions}
To summarize, we have used formal considerations for quantifying the strength of singlet-triplet coherence in radical-ion pairs, which is central for understanding their quantum state evolution. We have also applied the formalism of quantum retrodiction to provide a theoretically solid basis for deriving the master equation for radical-ion-pair quantum dynamics. This represents a refinement of our previous work, which is substantiated by Monte Carlo simulations. These have their own interest as they can realistically and precisely simulate the dynamics of RP reactions including all relevant physical processes. For most practical purposes, however, the master equation we derive should be adequate.

This work is about the self-consistency of our approach and not about making the case of which among the competing theories is the correct one.  In other words,  if the model presented in Fig. \ref{fig2} is a physically adequate model for describing RPs, as we believe it is, our newly introduced master equation represents a first-principles result alleviating problems with our previous phenomenological treatment. Nevertheless, we have compared the predictions of our approach with the other two competing theories and discussed in detail how all three theories describe the decay of S-T coherence, which is a central observable in RP reactions.
\appendix
\section{}
According to \cite{plenio_coherence}, any measure of coherence, $p_{\rm coh}(\rho)$, should satisfy the following requirements. \newline
(C1) $p_{\rm coh}(\rho)=0$ for $\rho\in\mathcal{I}$. \newline
(C2) $p_{\rm coh}(\rho)$ should be monotonous under all incoherent positive and trace-preserving maps, i.e. $p_{\rm coh}(\rho)\geq p_{\rm coh}(\Phi_{ICPTP}(\rho))$, where the map $\Phi_{ICPTP}(\rho)=\sum_{n}K_{n}\rho K_{n}^{\dagger}$ is defined by a set of Kraus operators $K_n$. These satisfy $\sum_{n}K_{n}^{\dagger}K_{n}=1$ and $K_{n}\mathcal{I}K_{n}^{\dagger}\subset\mathcal{I}$.\newline
(C3) There is a stronger requirement, namely that $p_{\rm coh}(\rho)$  is monotonous under selective measurements {\it on average}, namely $p_{\rm coh}(\rho)\geq\sum_{n} p_{n}p_{\rm coh}(\rho_n)$, where $\rho_n=K_n \rho K_{n}^{\dagger}/p_{n}$, again with $\sum_{n}K_{n}^{\dagger}K_{n}=1$ and $K_{n}\mathcal{I}K_{n}^{\dagger}\subset\mathcal{I}$. The probability to select $\rho_n$ in the measurement defined by $K_n$ is $p_n=\tr\{K_n\rho K_{n}^{\dagger}\}$.

We can now demonstrate that the previously defined measure \eqref{pcoh} is not a good measure of S-T coherence. An S-T decoherence process can be described by the following Kraus operators, $K_1=\sqrt{1-\lambda}{\rm {\rm Q_S}}$, $K_2=\sqrt{1-\lambda}{\rm Q_T}$ and $K_3=\sqrt{\lambda}\mathbbmtt{1}$. This set of operators has the effect of scaling $\rho_{\rm ST}$ and $\rho_{\rm TS}$ by the factor $0\leq\lambda\leq 1$. We would expect that the measure of coherence also scales by $\lambda$, however, defining $\rho'=\sum_{n=1}^{3}K_n \rho K_{n}^{\dagger}$ we easily find that $p_{\rm coh}(\rho')=\lambda^2 p_{\rm coh}(\rho)$ when using definition \eqref{pcoh} for $p_{\rm coh}$. Put differently, the measure \eqref{pcoh} is similar to the squared Hilbert-Schmidt norm $C_{l_2}(\rho)$, which does not satisfy \cite{plenio_coherence} the strong monotonicity criterion (C3). 
\section{}
To visualize the definition of ${\cal C}(\rho)$ in \eqref{crho} we consider a simple example of an S-T coherent state of a single-nucleus RP, e.g. $|\psi\rangle=\alpha|{\rm S}\rangle\otimes|\downarrow\rangle+\beta|{\rm T}_{-}\rangle\otimes|\uparrow\rangle$. The corresponding density matrix is
\begin{align}
\rho&=|\alpha|^2|{\rm S}\rangle\langle {\rm S}|\otimes|\downarrow\rangle\langle\downarrow|+|\beta|^2|{\rm T}_{-}\rangle\langle {\rm T}_{-}|\otimes|\uparrow\rangle\langle\uparrow|\nonumber\\
&+\alpha\beta^{*}|{\rm S}\rangle\langle {\rm T}_{-}|\otimes|\downarrow\rangle\langle\uparrow|+\alpha^{*}\beta|{\rm T}_{-}\rangle\langle {\rm S}|\otimes|\uparrow\rangle\langle\downarrow|\label{rho}
\end{align}
We wish to pick the amplitude $\alpha\beta^{*}$ of the S-T off-diagonal term in Eq. \eqref{rho}, i.e. the third term. This can be done as follows. In this simple example $\rho_{\rm ST}=\alpha\beta^{*}|{\rm S}\rangle\langle {\rm T}_{-}|\otimes|\downarrow\rangle\langle\uparrow|$. If we right-multiply $\rho_{\rm ST}$ with $|{\rm T}_{-}\rangle\langle {\rm S}|\otimes\mathbbmtt{1}_{2}$ we are left with $r=\alpha\beta^{*}|{\rm S}\rangle\langle {\rm S}|\otimes|\downarrow\rangle\langle\uparrow|$. If we then right-multiply $r$ with $r^{\dagger}$ and take the trace of the resulting expression it readily follows that $|\alpha\beta|=\sqrt{\tr\{rr^{\dagger}\}}$. In the general case we will have $|{\rm S}\rangle\langle {\rm T}_{0}|$, $|{\rm S}\rangle\langle {\rm T}_{+}|$ and $|{\rm S}\rangle\langle {\rm T}_{-}|$ coherences, hence the definition \eqref{crho}.
\acknowledgements
We would like to thank Prof. Claude Fabre for bringing to our attention the formalism of quantum retrodiction. We acknowledge support from the European Union's Seventh Framework Programme FP7-REGPOT-2012-2013-1 under grant agreement 316165 and the European Union (European Social Fund - ESF) and Greek national funds through the Operational Program "Education and Lifelong Learning" of the National Strategic Reference Framework (NSRF) - Research Funding Program THALIS.

\end{document}